# Charge and Field Driven Integrated Optical Modulators: Comparative Analysis

**JACOB B KHURGIN**[1*], **VOLKER J SORGER**[2], **RUBAB AMIN**[2]
[1]*Johns Hopkins University, Baltimore MD 21208, USA*
[2]*George Washington University, Washington DC 20052, USA*
*[jakek@jhu.edu](mailto:jakek@jhu.edu)

**Abstract: Electro optic modulators being key for many signal processing systems must adhere to requirements given by both electrical and optical constrains. Distinguishing between charge driven (CD) and field driven (FD) designs, we answer the question of whether fundamental performance benefits can be claimed of modulators based on emerging electro-optic materials. Following primary metrics, we compare the performance of emerging electro-optic and electro-absorption modulators such as graphene, transparent conductive oxides, and Si, based on charge injection with that of the 'legacy' FD modulators, such as those based on lithium niobate and quantum confined Stark effect. We show that for rather fundamental reasons, FD modulators always outperform CD ones in the conventional wavelength scale waveguides. However, for waveguide featuring a sub-wavelength optical mode, such as those assisted by plasmonics, the emerging CD devices are indeed highly competitive.**

Optical modulators are key components of any photonic system. Historically, for many years, electro-optic (i.e. based on refractive index change) modulators (EOM) made of $LiNbO_3$ (LN) have been devices-of-choice when it comes to data speed [1, 2]. All the attractive features of LN EOM– high speed, low insertion loss, decent efficiency, wide optical (spectral) bandwidth, absence of frequency chirp, come at the expense of a major drawback – the size. This is why for about 20 years many photonic integrated circuits (PIC) have been using electro-absorption modulators (EAM) based on quantum confined Stark effects (QCSE) in III-V quantum wells [3, 4]. QCSE (and similar Franz–Keldysh (FK) effect [5]) EAMs are more compact since they use thin epitaxial layers rather than diffused waveguides, while trading in the versatility of EOMs which are capable of supporting a variety of modulation formats. Thus, EOMs and EAMs have been co-existing each in its own niche up until the advent of silicon photonics, which demands modulators to be integrated on chip platform. This has been first achieved by integration of III-V devices on Si [6], and later by the development of thin film LN technology [7, 8]. But integration on Si platform requires additional fabrication steps, and, while the size of thin film LN modulators is reduced, further size reduction is required in next generation modulators [9, 10]. For this reason, the dominance of LN and QCSE modulators has been first challenged by Si modulators based on depletion/accumulation of carriers[11, 12], and in more recently by modulators based on emerging materials including transparent conductive oxides (TCO) , i.e. indium tin oxide ITO [13-16] and 2D materials[17, 18], such as graphene [19-21] and transition metal di-chalcogenides (TMDC) [22, 23]. In all these devices the modulating action is due to injection of carriers into the active material itself, rather than simply applying an electric field across the active layer. We shall refer to them as charge driven (CD), rather than field driven (FD) modulators. Often CD modulators are used in conjunction with plasmonic techniques that allows a tighter, subwavelength optical confinement.

There is a healthy number of literature on EOM and EAM modulators [24-27], yet it appears that there exists no direct and all-encompassing comparison of the performance of different materials and techniques. In fact, modulator performance depends not only on the active material and modulation technique (CD or FD), but also on geometry, electrical circuitry, and various extrinsic enhancement techniques, (microresonators, photonic crystals, etc.) Therefore, it is difficult to separate pure material factors as single stand-alone parameters, especially when comparing very different modulator setups. Building on our prior work [28, 29], we develop such single metrics, for both EOMs and EAMs to perform a holistic comparison. The conclusions are thought-provoking and unequivocal; FD modulators are superior when it comes to conventional waveguide dimension with thicknesses measured in 100's of nm, whereas in smaller (usually plasmonic-assisted waveguides), CD modulators' performance truly shines.

Two classes of modulators are shown in **Fig.1**; the CD modulator (**Fig.1a**) incorporates an active region and a control gate. The active region is shown schematically as a multilayer structure with $N \geq 1$ active layers, which may be monolayers of graphene, or other 2D material, semiconductor quantum wells (QWs), or layers of quantum dots (QDs). Even if the modulating material is three-dimensional, such as ITO, one can still use the 'layered picture'. For the sake of simplicity in this illustration we consider the two level entity, such as QDs or excitonic transition in TMDC. The absorption $\alpha$ and index $n$ in the absence of injected charge is highlighted in **Fig.1c**. When charge $Q$ is injected into the active region, the transition gets saturated. This leads to a reduction in both $\alpha$ at the peak wavelength $\lambda_0$ and $n$ at $\lambda_1 > \lambda_0$, appropriately chosen to avoid excessive absorption (**Fig.1d**). Thus, either amplitude modulation at $\lambda_0$ or phase modulation at $\lambda_1$ is achieved.

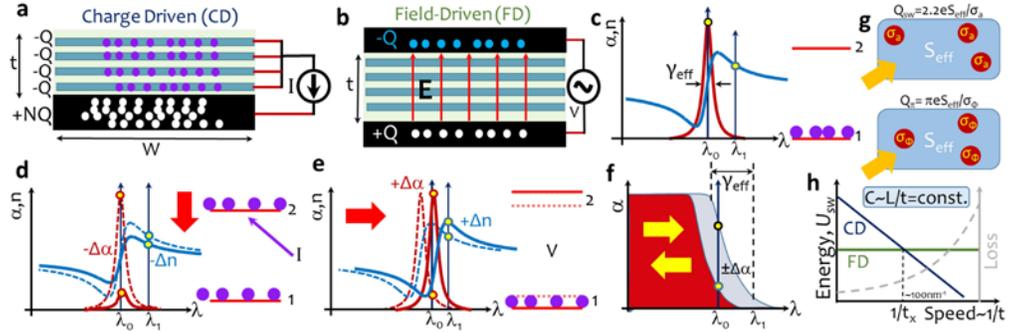

**Figure 1.** Comparison of fundamental principles of **(a)** electrical charge ($Q$)-driven (CD) modulators vs **(b)** Field ($E$)-driven (FD) design setup. **(c)** Absorption, $\alpha$ (red) and index, $n$ (blue) in the absence of charge/voltage in a two-level model. **(d)** Changes ($\Delta\alpha$ and $\Delta n$) imposed by injected charge in CD modulators; and **(e)** by applied field in FD modulator. **(f)** Changes imposed by charge or field as absorption edge shifts in the modulators with wide band absorption. **(g)** Intuitive picture of modulation – each injected carrier is represented as a perfect absorber 'disc' with area, $\sigma_a$ or a half-wave plate 'disc' with area, $\sigma_\Phi$ – when the effective area $S_{eff}$ is fully covered, full modulation is achieved. **(h)** Sketch of the effect of scaling down on switching energy, $U_{sw}$ for CD and FD modulators on a log-log scale. The ratio of length, $L$ to the thickness, $t$ (i.e. capacitance and modulation speed) is kept constant. The insertion loss is also presented.

In the FD modulator (**Fig.1b**) the charge appears only on two electrodes *outside the active region*, and the switching is performed by the field $E=Q/\varepsilon_0\varepsilon_s WL$ where $\varepsilon_s$ is static permittivity, $W$ is the width and $L$ is the length. A plane capacitor geometry is assumed, but in practical devices $W$ and $t$ should be understood as *effective* dimensions that take into account the fringe fields. In the absence of an applied voltage, i.e. plate charge $Q$, $\alpha$ a t wavelength $\lambda_0$ (below the peak) is small as shown in Fig.1c. When the charge is generated the absorption peak moves towards longer wavelength – Stark or FK effect (**Fig.1e**), increasing $\alpha$ at $\lambda_0$ and $n$ at $\lambda_1 > \lambda_0$.

In case of semiconductors and graphene, the band-to-band absorption is continuous, but the injection/depletion of carriers or Stark effect only change the absorption in a relatively narrow spectral region near the edge (**Fig.1f**). Hence the picture of **Fig.1(d,e)** remains valid as long as one is interested in changes of $\alpha$ and $n$. If one considers a case when free carriers are injected (e.g. ITO) this picture still holds as long as we assume the resonance frequency being 0.

The key parameter that defines the modulator performance is the total charge $Q_{sw}$ injected into the active layer in CD modulators or applied to the electrodes of FD modulators and required to attain switching. For amplitude-altering modulators (EAM), switching can be defined by a certain change in absorption, which in [28] was chosen to be 10dB. For the phase (EO) modulator switching occurs when the phase shift of 180 degrees is attained[29]. Using $Q_{sw}$ rather than $V_\pi$ makes it independent on the capacitance, $C$ which can be varied by altering geometry and permittivity. Then, the switching energy per bit is $U_{sw}=½Q_{sw}^2/C$, whereas the bandwidth is determined by $B\sim1/RC$, where $R$ is input impedance of the modulator matched to the impedance of the transmission line connecting the modulator to a driver (typically few tens of Ω's). Clearly, there always exists a tradeoff between $U_{sw}$ and $B$, but the ratio of two (or the product of $U_{sw}$ and switching time) $U_{sw}\tau_{sw}=½Q_{sw}^2R$ remains constant even as $C$ changes.. Thus, switching energy-time product $U_{sw}\tau_{sw}$ and by extension $Q_{sw}$ are the best fundamental figures of merit for any modulators that depend only on essential material properties and the degree of optical field confinement, as explained below.

We start with the CD EAM. As has been shown in [28] $Q_{sw}$ can be evaluated as $Q_{sw}=2.2eS_{eff}/\sigma_a$ which depends only on two parameters – the effective waveguide area, $S_{eff}$ and differential absorption cross-section of the material, $\sigma_a$ (**Fig.1g**). The effective area can be found as $S_{eff}=S_{act}/\Gamma$, where $S_{act}$ is the cross-section area of the active region, and $\Gamma = \iint_{active} E^2(x,y)\,dxdy/\iint_{\infty} E^2(x,y)\,dxdy$ is a confinement factor. The differential absorption cross-section of the material is the change of the absorption coefficient imposed by injection of a single electron (hole) and can be found in CD EAMs, independent of geometry, as[28] $\sigma_a(\omega)=A\pi\alpha_0\hbar/m_{eff}\gamma_{eff}$. Here, $A$ is a material-dependent coefficient that is always of the order of unity, $\alpha_0=1/137$ is a fine structure constant, $m_{eff}$ is the effective mass (for graphene one can introduce $m_{eff}(\omega)=2\hbar\omega/v_F^2$, where $v_F$ is Fermi velocity). $\gamma_{eff}$ is the effective broadening of the transition. For the QD or excitonic TMDC media $\gamma_{eff}$ is just the broadening of the material transition, , while for the state filling (Pauli-blocking) of Fig.1f the broadening is thermal $\hbar\gamma_{eff}=3k_BT$. The effective mass is inversely proportional to the bandgap, i.e. to the photon energy, and is pretty much fixed for a given value of $\hbar\omega$. As a result, for all the aforementioned mechanisms one obtains $\sigma_a\approx10^{-16}$cm$^2/\hbar\gamma_{eff}$, where $\hbar\gamma_{eff}$ is in meV[28]. The outcome is rather striking yet logical – $\sigma_a$ depends only on the value of the dipole matrix element of the transition and the broadening. For the allowed transition at a given frequency the value of that dipole is fixed by the oscillator sum rule and in the visible/near IR region is typically on the scale of bond length, i.e. a few Å[30]. *In the end, the strength of the absorption is not relevant. This is because the stronger the absorption (e.g. exciton vs. regular 2-level system), the more it is robust and hence more difficult to alter. These two factors balance each other out.* Note, that free carrier (e.g. Si, ITO) EAMs operating far from resonance have $\gamma_{eff}\sim\omega^2/\gamma$ i.e. is rather large and they would not be competitive, but for all other CD EAMs $\sigma_a\sim10^{-15}$cm$^2$.

For CD EOM modulation the switching charge required for the $\pi$ phase shift can be written as $Q_\pi=\pi eS_{eff}/\sigma_\phi$, where the phase shift cross-section $\sigma_\phi$ is the phase shift per one injected carrier. One can estimate (in line with KK relation) as $\sigma_a(\omega)=B\pi\alpha_0\hbar/m_{eff}\Delta\omega_{eff}$, where $B$ is on the order of unity, and $\Delta\omega_{eff}$ is effective detuning chosen to avoid spurious changes of absorption. For the QD and TMDC modulators $\Delta\omega_{eff}$ is the detuning from the absorption edge, and for ITO and graphene it is frequency itself, $\omega$. Note that since EOMs operate away from resonance, free carrier-based EOMs such as ITO are highly competitive [29]. As a consequence, and applicable for all the CD EOMs, we find $\sigma_\Phi\sim1–3\times10^{-16}$ cm$^2$ that the switching energy time product is about one order of magnitude higher that for EAMs, which maybe a small price to pay for the EOMs advantages, especially their lower insertion loss.

Now turn attention to FD EAM, such as, e.g. QCSE. The shift of the absorption edge is proportional to the electric field $E$ as $\Delta(\hbar\omega_{edge})\sim eE\Delta z_{cv}$, where $\Delta z$ is the effective dipole commensurate with the width of the QW[31]. Then the change of $\alpha$ is proportional to the above shift and the joint density per unit energy. In the end, the differential cross section for QCSE

FD EAM is $\sigma_{a,QCSE} \sim \pi\alpha_0 e^2 \Delta z N_{QW}/2\varepsilon_0\varepsilon_s\hbar\gamma_{eff}$, where $N_{QW}$ is the number of QWs. Exemplary, assuming an InP based modulator with 5 nm QWs ($\Delta z \sim 1$ nm) we obtain $\sigma_{a,QCSE} \sim 0.5\times10^{-16}$ cm$^2 \times N_{QW}/\hbar\gamma_{eff} \sim N_{QW}\sigma_a$. For FD modulator $\sigma_a$ is not a pure material parameter but also depends on the thickness of active region. The efficiency of the QCSE FD EAM increases with the number of QWs and can exceed the efficiency of CD EAM by an order of magnitude or more, as long as there is space for 10-20 QWs. Since the period is about 20 nm, such space can only be found in the relatively thick dielectric waveguides and not in the plasmonic ones, where CD modulators gain advantage. We can visualize this in Fig.1; *In FD modulators the field induced by the charge Q on electrodes penetrates the entire waveguide cross section causing the change in all available (N) layers (Fig. 1b), whereas in CD modulator (Fig.1a) the change occurs only in a single layer into which the same charge Q has been injected.* Hence the thicker is waveguide the more advantageous is FD modulator. However, once a good overlap between the optical mode and active medium ($\Gamma \sim 1$) is achieved, the reduction in $S_{eff}$ can only be attained by reducing $S_{act}$, which will also reduce $N_{QW}$ accordingly, and the $Q_{SW}$ will remain the same. In CD modulators, on the other hand, reduction in $S_{eff}$ always causes decrease in $Q_{SW}$ which gives them advantage in sub-$\lambda$ devices. This is sketched in **Fig.1(h)** –as one tries to scale down the modulator by reducing $L$ and $t$ (keeping $L/t$, i.e $C$ and modulation speed constant), $U_{sw}$ stays the same in FD devices, while loss increases. For the CD devices $U_{sw}$ decreases as $L^{-2} \sim t^{-2}$ which is extremely attractive, even taken higher loss in consideration. Below critical thickness $t_x \sim 100$nm for both QCSE and LN (see below) devices, the CD modulators gain clear advantage.

Next, we consider LN EOM, where the key parameter is Pockels coefficient, $r_{33}=30$ pm/V and static permittivity $\varepsilon_s=28$ and obtain $\sigma_{\phi,LN} = K\pi t_a e n^3 r_{33}/\varepsilon_0\varepsilon_s\lambda \approx 4\times10^{-18}$cm$^2\times t_a$, where $K\sim1$ is related to polarization and $t_a$ is the active layer thickness in nm. Again, we find that in this FD modulator $\sigma_\phi$ increases with the thickness since each carrier placed on electrode changes refractive index in the entire active layer. For a waveguide with a 500 nm active layer thickness $\sigma_{\phi,LN} \sim 0.5 \times 10^{-15}$cm$^2$, i.e. about an order of magnitude higher than in any CD EOM. This result is a bit surprising considering that the LN EOM operates far away from the absorption edge, but the nature of EO effect in LN is different from that effect in QWs. The origin of the Pockels effect in LN is the very strong change of the bond lengths [32, 33].

In conclusion, we have set out to provide an answer on how modulator performance is fundamentally impacted between electric field-driven (FD) versus charge driven (CD) setups. In both cases, electrical charge must move – either to change the carrier concentrations directly (CD) or via loading a capacitor (FD). The required switching charge $Q_{sw}$ or $Q_\pi$ forms the fundamental metric for all modulators. The main difference between the FD and CD modulators is that once the charge is introduced into the electrodes, the resulting absorption and index modulation is 'felt' throughout the entire waveguide in the FD case, whereas it is only modulated locally in the CD case. We find that the legacy FD devices, both EAM and EOM, will hold an advantage for larger waveguides of a few hundred nm thickness. But if one attempts to further reduce switching charge by reducing effective area, employing ultra-thin, plasmonic, or hybrid waveguides the advantages of FD fade and CD modulators, such as ITO and graphene among others, become a very attractive alternative for either index or absorption modulation. Another thought-provoking observation is that from the fundamental material point of view the change of absorption (index) per unit injected charge, all CD materials are roughly comparable and there is no reason to expect a new material with dramatically improved performance to appear on the scene. So, all future improvements will in all probability come not as much from fundamental material breakthroughs as from further gradual progress in miniaturization while lowering insertion loss, series resistance and refining low cost fabrication techniques.